# Popper's Experiment and the Uncertainty Principle

A. Cardoso

*Centro de Filosofia das Ciências da Universidade de Lisboa, Faculdade de Ciências da Universidade de Lisboa, Campo Grande, Edifício C4, 1749-016 Lisboa, Portugal*

Abstract: In this paper we look at a particular realization of Popper's thought experiment with correlated quantum particles and argue that, from the point of view of a nonlinear quantum physics and contrary to the orthodox interpretation, Heisenberg's uncertainty principle is violated. Moreover, we show that this kind of experiments can easily be explained in an intuitive manner if we are willing to take a nonlinear approach.

## 1. Introduction

Almost a century ago Karl Popper proposed an experiment[1] that could test the validity of the orthodox interpretation of quantum mechanics[2]. His proposal is very similar, although much less know, that the popular EPR thought experiment[3]. His early proposal, made in 1934, was later reformulated and finally published in 1983.

Popper proposed a setup where two quantum particles correlated in space and time, resulting from a common point source, are emitted in opposite directions towards two different detectors. Due to momentum conservation of the particle pair, if we detect one particle coming in a certain direction then we know that its twin is travelling in the opposite direction. Now, before each detector a narrow slit is placed so that the momentum of each particle is widely scattered. The slits are adjusted to have the same width, in which case, according to the orthodox or any other interpretation, the momenta of the particles are expected to show equal scatter. Then, and more interestingly, one of the slits is removed so that we learn the position of the unconstrained particle only by knowing that its twin has gone through the remaining slit. The question is now whether the mere knowledge of the position of a particle would be enough to increase the scatter, and therefore the uncertainty, in its momentum. Popper argued that according to the orthodox view the answer is yes, and from his realistic point of view the answer is no.

In this paper we will focus on a particular realization of Popper's experiment performed in 1999 by Kim and Shi[4], who conclude from their results that, as predicted by Popper, a constraint on the position of one particle does not imply a spread in the momentum of another particle entangled with the first. Moreover, they tell us that in their experiment Heisenberg's uncertainty principle – which states that the simultaneous prediction of the position and momentum, with uncertainties $\Delta x$ and $\Delta p$, respectively, of a quantum particle is limited by the relation $\Delta x \Delta p \geq h$, where $h$ is Planck's constant – is apparently violated.

The authors then argue, however, that their results are not in contradiction with the orthodox approach, as in a system of two entangled particles each individual particle cannot be viewed as an independent entity. Thus, the uncertainty principle in the form mentioned above does not apply in this case. This argument was later discussed and reinforced by Qureshi[5], who has shown that the results obtained in this experiment for the uncertainty of position and momentum are the expected ones for a system of two entangled particles.

As shown in a recent paper[6], the currently accepted interpretation of quantum mechanics has been questioned by the results of a new double-slit experiment using correlated photon pairs[7]. In particular, Niels Bohr's complementarity principle, an essential feature of the orthodox approach, has been clearly violated. On the other hand, those same results are perfectly understandable from the point of view of a nonlinear quantum physics[8] inspired in the early ideas of Louis de Broglie[9]. Therefore, it is only natural that one may want to look at and reinterpret the results presented by Kim and Shih but now taking a nonlinear approach.

We will first present Kim and Shih's realization of Popper's thought experiment and show that their results can be easily and intuitively understood if we take a nonlinear approach. We note that this same approach can be used to explain the results from other, similar experiments – the so-called 'ghost' diffraction and interference experiments – performed by different authors[10]. We will then argue that, even though from an orthodox point of view two particles correlated in space and time cannot be seen as individual entities, in the framework of a nonlinear approach those particles are independent despite having a common origin, and thus in this realization of Popper's thought experiment the uncertainty principle is in fact violated.

## 2. A realization of Popper's experiment

The first part of Kim and Shih's realization of Popper's experiment can be presented in a simplified manner using the setup shown in Fig. 1.

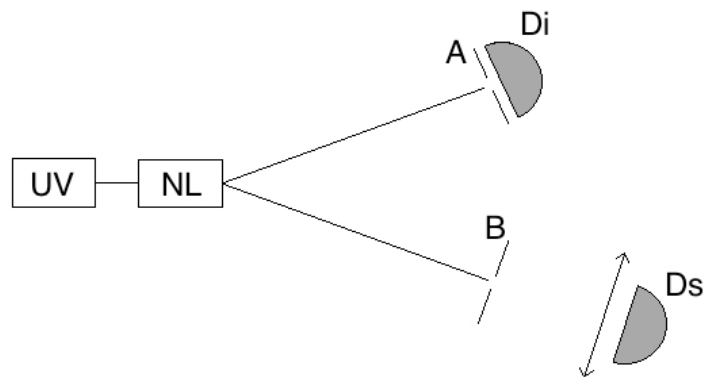

Fig. 1 – Kim and Shih's realization of Popper's thought experiment with both slits present.

A UV pump beam is injected onto a nonlinear crystal NL, which transforms an incoming photon $\psi$ into a pair of photons – one called idler and represented by the wave-function $\psi_i$ and the other one called signal, $\psi_s$. Only one pair of photons, correlated in space and time due to conservation of momentum, is produced at a time.

The idler photon $\psi_i$ is directed onto a single slit A and then incident onto detector $D_i$ fixed just behind the slit, whereas the signal photon $\psi_s$ is incident onto slit B – whose width is the same as the width of slit A – and its wave diffracts before reaching detector $D_s$, which is scanned along a direction perpendicular to the photon's trajectory. The signal at $D_s$ can then be counted alone or in coincidence with the detections at $D_i$.

In this situation a diffraction pattern is obtained at $D_s$, with a width that does not depend on whether the detections are done alone or in coincidence with the counts at $D_i$.

In the second part of the experiment (see Fig. 2) slit B is removed, and thus the signal photon $\psi_s$ is directly incident onto detector $D_s$ while the corresponding idler photon still $\psi_s$ goes through slit A before reaching detector $D_i$.

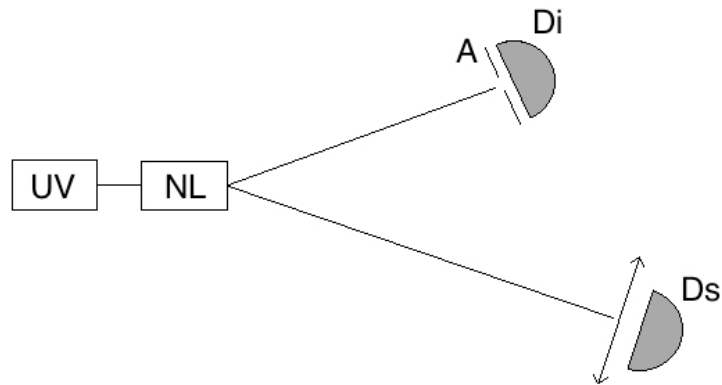

Fig. 2 – Kim and Shih's realization of Popper's thought experiment with one slit removed.

In this case a pattern similar to the one obtained previously but with a smaller width is observed at detector $D_s$ when the counts are done in coincidence with $D_i$. The counting rate at $D_s$ alone, however, is now constant in the scanning range.

## 2.1. Non-linear interpretation of the results

From the point of view of a nonlinear quantum physics the photon, or any quantum particle, is composed of an extended real part – the guiding wave – plus a localized corpuscle. This physical wave guides the corpuscle preferentially to the regions were the intensity of the wave is greater. As the photon arrives at a detector only the corpuscle, which carries almost all the energy of the particle, is observed, whereas its guiding wave interacts with the detector but doesn't have enough energy to trigger it.

Now, in the first part of Kim and Shih's experiment, the guiding wave $\psi_s$ of any signal photon passing through slit B will diffract before reaching detector $D_s$. Thus, because the corresponding corpuscle preferentially follows a path where the intensity of its wave is high, a diffraction pattern is observed as the detections of signal photons accumulate on the screen. Naturally, as the slits have the same width, the idler photons $\psi_i$ that pass through slit A correspond to the signal ones that went through slit B, and thus the signal at $D_s$ does not depend on whether the counts have been made alone or in coincidence with the detections at $D_i$.

In this case, as already discussed by the authors from the point of view of the orthodox interpretation, the uncertainty in position $\Delta x_s$ of the signal photon is equal to the width of slits A and B, and thus uncertainty in its momentum $\Delta p_s$ is such that $\Delta x_s \Delta p_s \approx h$[11].

In the second part of the experiment, only the idler photons that have passed through slit A will reach detector $D_i$, and thus, due to the spatial correlation between the idler-signal pair, only the corresponding signal photons will be detected in coincidence at $D_s$. Now, because there is no slit in the path of the signal photon, its wave $\psi_s$ will not be diffracted and thus the pattern observed at $D_s$ when the counts are done in coincidence with the detections at $D_i$ will be narrower than the diffraction pattern observed when slit B is present. Moreover, it is natural that in this case the detections at $D_s$ alone show a pattern which is constant in the scanning range, as all the photons in the diverging signal beam reach the detector, including the ones corresponding to idler photons that do not go through slit A and thus do not reach $D_i$.

In this situation, due to the spatial correlation between the photon pair, the uncertainty in position $\Delta x_s$ of the signal photon is still equal to the width of slit A but the uncertainty in its momentum $\Delta p_s$ is now lower than in the previous case due to the narrowing of the pattern at detector $D_s$, and so we obtain $\Delta x_s \Delta p_s < h$.

As shown in a recent paper dedicated to quantum-eraser experiments[8], from the point of view of a nonlinear quantum physics the apparent interaction between the particle pair produced at the nonlinear crystal NL is just a detector-selection effect where an appropriate part of the signal is chosen. In Kim and Shih's experiment it is clear that the pattern obtained at detector $D_s$ when the counts are done in coincidence with $D_i$ is only part of a total signal created by the divergence of the signal beam.

Thus, according to this approach and contrary to the orthodox view – which, as mentioned above, has been questioned by new double-slit experiments using correlated photon pairs – the signal and idler photons are independent entities once they are created even though they are correlated due to momentum conservation. Therefore, we can conclude that in this experiment the uncertainty principle has been violated.

## 3. Conclusions

In this paper we have shown that Kim and Shih's realization of Popper's thought experiment can be intuitively understood in the framework of a nonlinear quantum physics. Moreover, the explanation presented here is valid as well for other, similar experiments like the ghost diffraction and interference experiments performed by several authors.

Finally, we have argued that, contrary to the conclusions taken in the framework of the orthodox interpretation of quantum mechanics, according to our nonlinear approach – because each particle in a correlated pair is independent once it is created – Heisenberg's uncertainty principle is actually violated in this experiment.